\documentclass[a4paper,onecolumn,11pt,unpublished,noarxiv,allowtoday]{quantumarticle}
\pdfoutput=1
\usepackage[utf8]{inputenc}
\usepackage[english]{babel}
\usepackage[T1]{fontenc}
\usepackage{amsmath}
\usepackage{hyperref}
\usepackage{mathtools, nccmath}
\usepackage{setspace}
\usepackage{hyperref}
\usepackage{outlines}
\usepackage{fancyvrb}
\usepackage{mathtools}
\usepackage{physics}
\usepackage{comment}
\usepackage{bbold}
\usepackage{amssymb}
\usepackage{cancel}
\usepackage{graphicx}
\usepackage[caption=false]{subfig}
\usepackage[usenames, dvipsnames]{color}
\usepackage{soul}
\usepackage{tensor}
\usepackage{amsthm}
\usepackage[caption=false]{subfig}
\usepackage{algorithm}
\usepackage{algcompatible}

\usepackage{tikz}
\usepackage{lipsum}
\usepackage{tkz-graph}
\usetikzlibrary{hobby,backgrounds,calc,trees}
\usepackage{xcolor}
\usepackage{soul}
\usetikzlibrary{positioning}
\usepackage{ulem}
\usepackage{yquant}
\usepackage{caption}
\usepackage{subfig}
\useyquantlanguage{groups}
\usepackage{makecell}
\usepackage{tikz,esvect,tikz-3dplot}
\usetikzlibrary{3d,calc,intersections}
\usepackage{tikz}
\usepackage{tikz-3dplot}
\usepackage{adjustbox}
\usepackage{zx-calculus}
\usetikzlibrary{zx-calculus}
\usepackage{tikz-cd}
\usepackage[official]{eurosym}

\makeatletter

\usepackage{tensor}
\usepackage{accents}
\usetikzlibrary{angles,quotes,fit}

\definecolor{zx_red}{RGB}{232, 165, 165}
\definecolor{zx_green}{RGB}{216, 248, 216}

\usepackage[numbers,sort&compress]{natbib}
\begin{document}

\title{Pauli webs spun by transversal $|Y\rangle$ state initialisation}

\date{\today}

\author{Kwok Ho Wan}
\email{kwok.wan14((at))imperial((dot))ac((dot))uk}
\orcid{0000-0002-1762-1001}
\affiliation{Blackett Laboratory, Imperial College London, South Kensington, London SW7 2AZ, UK}
\affiliation{Mathematical Institute, University of Oxford, Andrew Wiles Building, Woodstock Road, Oxford OX2 6GG, UK}

\author{Zhenghao Zhong}
\orcid{0000-0001-5159-1013}
\affiliation{Mathematical Institute, University of Oxford, Andrew Wiles Building, Woodstock Road, Oxford OX2 6GG, UK}
\affiliation{Blackett Laboratory, Imperial College London, South Kensington, London SW7 2AZ, UK}

\begin{abstract}
Originally motivated by the (fold-)transversal related initialisation of logical surface code $|Y\rangle$ states from \cite{Moussa_2016,Gidney_inplace_2024,gidney2023cleanermagicstateshook}, which was then explicitly extended to the fold-transversal $S$ gate implementation in \cite{chen2024transversallogicalcliffordgates} for the rotated surface code, we employ ZX-calculus and Pauli web to understand the $|Y\rangle=S|+\rangle$ state transversal initialisation scheme.
\end{abstract}

\maketitle

\section{Introduction}
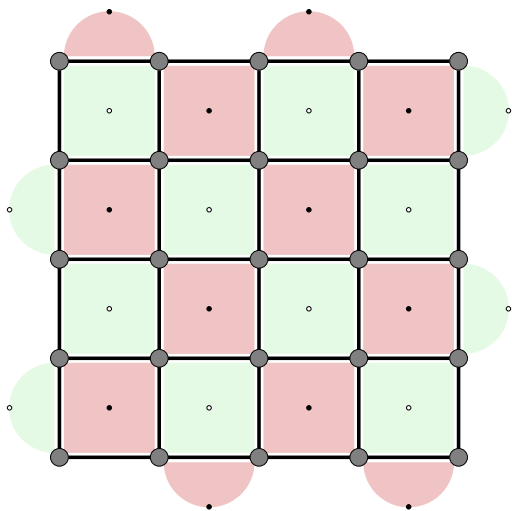
\begin{figure}[!h]
    \centering
    \resizebox{0.45\linewidth}{!}{
    \begin{tikzpicture}[scale=.7,every node/.style={minimum size=0.5cm},on grid]

    \node[fill=gray,shape=circle,draw=black] (n4) at (2,18) {};
    \node[fill=gray,shape=circle,draw=black] (n3) at (2,14) {};
    \node[fill=gray,shape=circle,draw=black] (n2) at (2,10) {};
    \node[fill=gray,shape=circle,draw=black] (n1) at (2,6) {};
    \node[fill=gray,shape=circle,draw=black] (n0) at (2,2) {};

    \node[fill=gray,shape=circle,draw=black] (n9) at (6,18) {};
    \node[fill=gray,shape=circle,draw=black] (n8) at (6,14) {};
    \node[fill=gray,shape=circle,draw=black] (n7) at (6,10) {};
    \node[fill=gray,shape=circle,draw=black] (n6) at (6,6) {};
    \node[fill=gray,shape=circle,draw=black] (n5) at (6,2) {}; 

    \node[fill=gray,shape=circle,draw=black] (n14) at (10,18) {};
    \node[fill=gray,shape=circle,draw=black] (n13) at (10,14) {};
    \node[fill=gray,shape=circle,draw=black] (n12) at (10,10) {};
    \node[fill=gray,shape=circle,draw=black] (n11) at (10,6) {};
    \node[fill=gray,shape=circle,draw=black] (n10) at (10,2) {};

    \node[fill=gray,shape=circle,draw=black] (n19) at (14,18) {};
    \node[fill=gray,shape=circle,draw=black] (n18) at (14,14) {};
    \node[fill=gray,shape=circle,draw=black] (n17) at (14,10) {};
    \node[fill=gray,shape=circle,draw=black] (n16) at (14,6) {};
    \node[fill=gray,shape=circle,draw=black] (n15) at (14,2) {}; 

    \node[fill=gray,shape=circle,draw=black] (n24) at (18,18) {};
    \node[fill=gray,shape=circle,draw=black] (n23) at (18,14) {};
    \node[fill=gray,shape=circle,draw=black] (n22) at (18,10) {};
    \node[fill=gray,shape=circle,draw=black] (n21) at (18,6) {};
    \node[fill=gray,shape=circle,draw=black] (n20) at (18,2) {};

    \begin{pgfonlayer}{background}

    \draw[zx_green,fill=zx_green,opacity=0.65](1.8,2.2) to[curve through={(0,4)}](1.8,5.8);
    \filldraw[fill=zx_red, opacity=0.65, draw=zx_red] (2.2,2.2) rectangle (5.8,5.8);
    \filldraw[fill=zx_green, opacity=0.65, draw=zx_green] (2.2,6.2) rectangle (5.8,9.8);
    \filldraw[fill=zx_red, opacity=0.65, draw=zx_red] (2.2,10.2) rectangle (5.8,13.8);
    \draw[zx_green,fill=zx_green,opacity=0.65](1.8,10.2) to[curve through={(0,12)}](1.8,13.8);
    \filldraw[fill=zx_green, opacity=0.65, draw=zx_green] (2.2,14.2) rectangle (5.8,17.8);
    \draw[zx_red,fill=zx_red,opacity=0.65](2.2,18.2) to[curve through={(4,20)}](5.8,18.2);

    \draw[zx_red,fill=zx_red,opacity=0.65](9.8,1.8) to[curve through={(8,0)}](6.2,1.8);
    \filldraw[fill=zx_green, opacity=0.65, draw=zx_green] (6.2,2.2) rectangle (9.8,5.8);
    \filldraw[fill=zx_red, opacity=0.65, draw=zx_red] (6.2,6.2) rectangle (9.8,9.8);
    \filldraw[fill=zx_green, opacity=0.65, draw=zx_green] (6.2,10.2) rectangle (9.8,13.8);
    \filldraw[fill=zx_red, opacity=0.65, draw=zx_red] (6.2,14.2) rectangle (9.8,17.8);

    \filldraw[fill=zx_red, opacity=0.65, draw=zx_red] (10.2,2.2) rectangle (13.8,5.8);
    \filldraw[fill=zx_green, opacity=0.65, draw=zx_green] (10.2,6.2) rectangle (13.8,9.8);
    \filldraw[fill=zx_red, opacity=0.65, draw=zx_red] (10.2,10.2) rectangle (13.8,13.8);
    \filldraw[fill=zx_green, opacity=0.65, draw=zx_green] (10.2,14.2) rectangle (13.8,17.8);
    \draw[zx_red,fill=zx_red,opacity=0.65](10.2,18.2) to[curve through={(12,20)}](13.8,18.2);
    
    \draw[zx_red,fill=zx_red,opacity=0.65](17.8,1.8) to[curve through={(16,0)}](14.2,1.8);
    \filldraw[fill=zx_green, opacity=0.65, draw=zx_green] (14.2,2.2) rectangle (17.8,5.8);
    \filldraw[fill=zx_red, opacity=0.65, draw=zx_red] (14.2,6.2) rectangle (17.8,9.8);
    \draw[zx_green,fill=zx_green,opacity=0.65](18.2,9.8) to[curve through={(20,8)}](18.2,6.2);
    \filldraw[fill=zx_green, opacity=0.65, draw=zx_green] (14.2,10.2) rectangle (17.8,13.8);
    \filldraw[fill=zx_red, opacity=0.65, draw=zx_red] (14.2,14.2) rectangle (17.8,17.8);
    \draw[zx_green,fill=zx_green,opacity=0.65](18.2,17.8) to[curve through={(20,16)}](18.2,14.2);
    \end{pgfonlayer}

    \begin{pgfonlayer}{background}
        \path [-,line width=0.1cm,black,opacity=1] (n0) edge node {} (n20);
        \path [-,line width=0.1cm,black,opacity=1] (n1) edge node {} (n21);
        \path [-,line width=0.1cm,black,opacity=1] (n2) edge node {} (n22);
        \path [-,line width=0.1cm,black,opacity=1] (n3) edge node {} (n23);
        \path [-,line width=0.1cm,black,opacity=1] (n4) edge node {} (n24);

        \path [-,line width=0.1cm,black,opacity=1] (n0) edge node {} (n4);
        \path [-,line width=0.1cm,black,opacity=1] (n5) edge node {} (n9);
        \path [-,line width=0.1cm,black,opacity=1] (n10) edge node {} (n14);
        \path [-,line width=0.1cm,black,opacity=1] (n15) edge node {} (n19);
        \path [-,line width=0.1cm,black,opacity=1] (n20) edge node {} (n24);

        \node[fill=black,circle,draw=black,scale =0.25] (b0) at (4,4) {};
        \node[fill=black,circle,draw=black,scale =0.25] (b1) at (4,12) {};
        \node[fill=black,circle,draw=black,scale = 0.25] (b2) at (4,20) {};

        \node[fill=black,circle,draw=black,scale=0.25] (b3) at (8,0) {};
        \node[fill=black,circle,draw=black,scale = 0.25] (b4) at (8,8) {};
        \node[fill=black,circle,draw=black,scale = 0.25] (b5) at (8,16) {};

        \node[fill=black,circle,draw=black,scale=0.25] (b6) at (12,4) {};
        \node[fill=black,circle,draw=black,scale = 0.25] (b7) at (12,12) {};
        \node[fill=black,circle,draw=black,scale = 0.25] (b8) at (12,20) {};

        \node[fill=black,circle,draw=black,scale=0.25] (b9) at (16,0) {};
        \node[fill=black,circle,draw=black,scale = 0.25] (b10) at (16,8) {};
        \node[fill=black,circle,draw=black,scale = 0.25] (b11) at (16,16) {};

        \node[fill=white,circle,draw=black,scale = 0.25] (a0) at (0,4) {};
        \node[fill=white,circle,draw=black,scale = 0.25] (a1) at (0,12) {};

        \node[fill=white,circle,draw=black,scale = 0.25] (a2) at (4,8) {};
        \node[fill=white,circle,draw=black,scale = 0.25] (a3) at (4,16) {};

        \node[fill=white,circle,draw=black,scale = 0.25] (a4) at (8,4) {};
        \node[fill=white,circle,draw=black,scale = 0.25] (a5) at (8,12) {};

        \node[fill=white,circle,draw=black,scale = 0.25] (a6) at (12,8) {};
        \node[fill=white,circle,draw=black,scale=0.25] (a7) at (12,16) {};

        \node[fill=white,circle,draw=black,scale = 0.25] (a8) at (16,4) {};
        \node[fill=white,circle,draw=black,scale=0.25] (a9) at (16,12) {};

        \node[fill=white,circle,draw=black,scale=0.25] (a10) at (20,8) {};
        \node[fill=white,circle,draw=black,scale=0.25] (a11) at (20,16) {};

    \end{pgfonlayer}

\end{tikzpicture}
    }
    \caption{A distance $d=5$ rotated surface code with green $X$-type and red $Z$-type plaquettes. The larger $d^2$ gray nodes represents the data qubits whilst the smaller $d^2-1$ black and white nodes are the syndrome qubits initialised in the $\ket{0}$ and $\ket{+}$ states respectively for syndrome extraction.}
    \label{fig:surface_code}
\end{figure}

The surface code allows fault-tolerant transversal logical $\ket{0}$ and $\ket{+}$ state initialisations \cite{Fowler_2012}. This can be carried out by initialising all the data qubits (gray nodes in figure \ref{fig:surface_code}) in the $\ket{0}$ or $\ket{+}$ states respectively and then performing multiple full rounds of parity measurement to end up with logical encoded $\ket{0}$ or $\ket{+}$ states. However, the remaining Pauli $Y$ eigenstate ($\ket{Y} \propto \ket{0}+i\ket{1}$), does not have simple transversal logical initialisations on the surface code. Many efforts had been made to overcome this, they can be broadly categorised into two approaches. Firstly, the native state injection schemes \cite{Li_2015,lao_criger_magic,gidney2023cleanermagicstateshook}, which probabilistically injects a physical to a logical $\ket{Y}$ state. Secondly, the fold-transversal $S$ gate implementation \cite{Moussa_2016,Gidney_inplace_2024,chen2024transversallogicalcliffordgates}, whereby non-local interaction (or the equivalent of using additional gates) within the surface code permits the implementation of a logical $S = \begin{pmatrix}
   1  & 0 \\
   0  & i
\end{pmatrix}$ gate. The $S$ gate in-turn can be applied to the transversally initialised logical $\ket{+}$ state, resulting in the $\ket{Y}$ state (as $\ket{Y}=S\ket{+}$). We shall provide a brief summary on these different schemes.
\begin{enumerate}
   \item Moussa's fold-transversal $S$ gate \cite{Moussa_2016} can be applied to the transversally initialised logical $\ket{+}$ state on the un-rotated surface code to generate a logical $\ket{Y}$ state. Unfortunately, for the rotated surface code, Moussa's scheme will require the elongation of the rotated surface code in order to perform a fold-transversal $S$ gate.
\item Li, Lao-Criger, Gidney's related state injection schemes \cite{Li_2015,lao_criger_magic,gidney2023cleanermagicstateshook} can be applied to inject a bare physical $\ket{Y}$ state to encode a surface code logical $\ket{Y}$ state probabilistically.  
 \item Gidney's inplace $Y$ eigenstate initialisation scheme can generate a logical $\ket{Y}$ state in $\lfloor \frac{d}{2} \rfloor+2$ code cycles for a distance $d$ rotated surface code \cite{Gidney_inplace_2024}\footnote{Note that Gidney also devised a modified fold-transversal $S$ gate for the rotated surface code availiable in his code upload on Zenodo \cite{Gidney_Zenodo_Inplace}. However, this will still require minor elongation of the rotated surface code.}.
 \item Chen-Chen-Lu-Pan's (CCLP) fold-transversal $S$ gate, which is somewhat akin to a hybrid of Moussa's fold-transversal \cite{Moussa_2016} and Gidney's schemes \cite{gidney2023cleanermagicstateshook,Gidney_inplace_2024}. The advantage of this $S$ gate utilises the decoding techniques from \cite{sahay2024errorcorrectiontransversalcnot,wan2024iterativetransversalcnotdecoder,wan2024constanttimemagicstatedistillation} and shows promising performance, implementing a logical $S$ gate in $\mathcal{O}(1)$, constant code cycles. 
\end{enumerate}

Originally, our calculations were performed on the less efficient $\ket{Y}$ state (fold-)/transversal initialisation schemes \cite{Moussa_2016,Gidney_inplace_2024,gidney2023cleanermagicstateshook}. This short paper is the re-iteration of the our original calculation applied to the CCLP scheme \cite{chen2024transversallogicalcliffordgates}. We shall illustrate their protocol using the language of ZX-calculus \cite{Coecke_2011} and Pauli web \cite{Bombin_2024,rodatz2024floquetifyingstabilisercodesdistancepreserving}. We hope that the  graphical calculus employed can provide an introduction to understanding complicated Clifford quantum states \cite{gottesman1997stabilizercodesquantumerror} as outlined in the CCLP fold-transversal $S$ gate scheme \cite{chen2024transversallogicalcliffordgates}.

The main results of this short paper is to verify that a logical $X$ correlator maps to the logical $Y$ correlator \cite{Bombin_2024} via the CCLP fold-transversal $S$ gate. This can be seen in its Pauli webs. We leave the code stabilisers for the readers to verify.

\section{Review on rotated surface code ZX-diagrams}
\begin{figure}[!h]
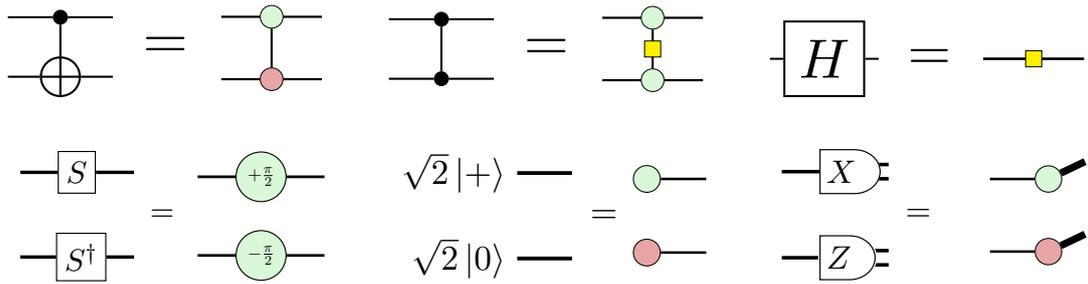

\centering
\resizebox{0.3\linewidth}{!}{

}
\caption{Here are some useful gates (CNOT, CZ, Hadamard, $S$/$S^{\dagger}$ phase gate) initialisations (initial states: $\ket{+}$ and $\ket{0}$) and measurements in the $X$ or $Z$ basis in the ZX-calculus notation.}
\label{fig:useful_zx_notations}
\end{figure}

The ZX calculus is a useful diagrammatic tool for representing and manipulating quantum systems \cite{Coecke_2011,vandewetering2020zxcalculusworkingquantumcomputer}. Prior to describing the CCLP fold-transversal $S$ gate implementation on the rotated surface code, we have listed all the most commonly used gates in the ZX calculus notation in figure \ref{fig:useful_zx_notations}. In addition, Pauli webs are a graphical notation system that can be drawn on top of Clifford ZX-diagrams concerning quantum error correcting codes to visualise and identify their checks, stabilizers and logical correlators. Please refer to \cite{rodatz2024floquetifyingstabilisercodesdistancepreserving, Bombin_2024,wan2025pauliwebyranglestate} for discussions on Pauli webs. They are extremely useful when verifying complicated Clifford circuits \cite{vandewetering2020zxcalculusworkingquantumcomputer,wan2025pauliwebyranglestate}.

\begin{figure}[!h]
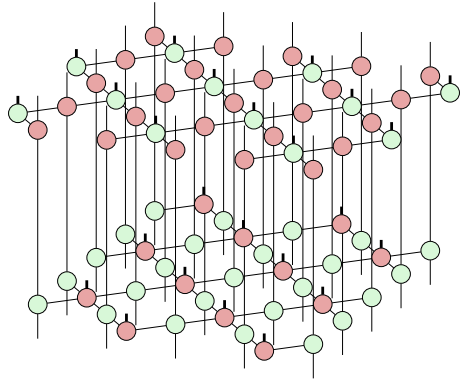

    \centering     
    \tdplotsetmaincoords{70}{23}
    \resizebox{0.4\linewidth}{!}{
   
    %
    }
    \caption{\label{fig:inj_encoder}A distance $d=5$ rotated surface code encoder circuit. This consists of Z-type followed by X-type plaquette parity measurements (for detailed discussion, see \cite{Fowler_2012,Bombin_2024}).}
\end{figure}

The surface code can be represented as a ZX-diagram. In figure \ref{fig:inj_encoder}, we show the logical encoder circuit for the surface code. This involves measuring all the Z- and X-type parity measurements associated with the stabilisers of the surface code \cite{Fowler_2012}. Time goes from bottom to top throughout all ZX-diagrams in this manuscript.  

\begin{figure}[!h]
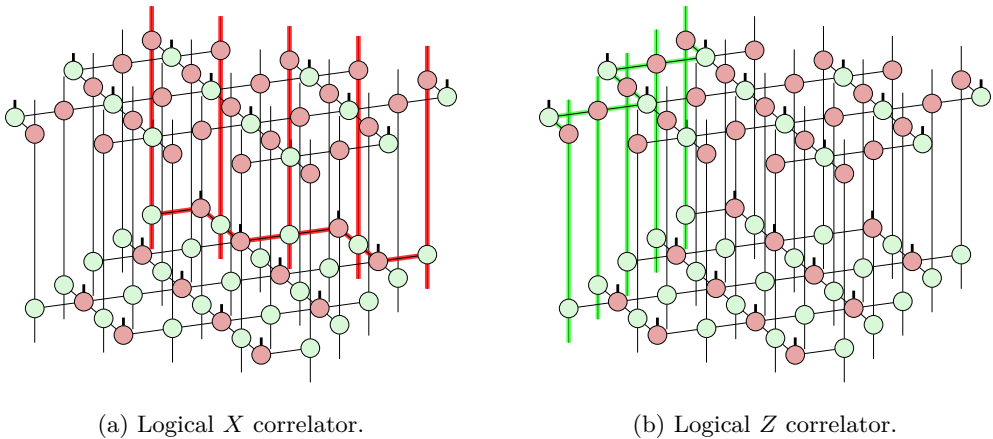

    \centering     
    \tdplotsetmaincoords{70}{23}

    \subfloat[][\label{fig:logical_cor_X}Logical $X$ correlator.]{
    \resizebox{0.4\linewidth}{!}{
   
    %
    }
    }
    \caption{\label{fig:logical_cor}These are the logical $X$ and $Z$ correlators, linking the logical ($X$ and $Z$ respectively) operators of the surface code from the input (bottom legs) to the output (top legs).}
\end{figure}
The surface code encoder circuit in figure \ref{fig:inj_encoder} have logical $X$ or $Z$ correlators \cite{Bombin_2024} that connects the $X$ or $Z$ logical operators from the input to output of the ZX-diagram (bottom to top) in figure \ref{fig:logical_cor}. The logical $Z$ ($X$) correlator is the green (red) Pauli web in figure \ref{fig:logical_cor_Z} (\ref{fig:logical_cor_X})\footnote{The $X$ and $Z$ surface code stabilisers are in the opposite orientation relative to our previous paper \cite{wan2025pauliwebyranglestate} to coincide with the orientation in \cite{chen2024transversallogicalcliffordgates}.}.

\section{CCLP fold-transversal $S$ gate}
We take the CCLP fold-transversal $S$ gate implementation circuit from their FIG. S1 (b)\footnote{Arxiv version 1.} \cite{chen2024transversallogicalcliffordgates}. We modify this circuit and initialise all the input data qubits in the $\ket{+}$ state, and re-write it as a ZX-diagram in figure \ref{fig:CCLP_no_web0} to figure \ref{fig:CCLP_no_web6}. By applying the CCLP fold-transversal $S$ gate circuit onto $\ket{+}$ initialised data qubits, we expect the final state to be a rotated surface code encoded logical $Y$ eigenstate. For illustration purposes, all the rotated surface code will be drawn to distance $d = 5$ in this manuscript. There are no reasons why the graphical ZX-diagram treatment outlined in this manuscript cannot be extended to higher than distance $5$ surface codes.

\begin{figure}[!h]
\centering
\subfloat[\label{fig:CCLP_no_web0}Time slice 0: the `\textit{post-reset state}' from \cite{chen2024transversallogicalcliffordgates}. Our only departure is to initialise all data qubits in the $\ket{+}$ state, while all the syndrome qubits will still be initialised in their respective $\ket{+}$ or $\ket{0}$ states.]{
\resizebox{0.4\linewidth}{!}{
    \centering     
    \tdplotsetmaincoords{65}{22.5}


}
}
\qquad
\subfloat[\label{fig:CCLP_no_web1}Time slice 1: the `\textit{waxing-crescent state}' from \cite{chen2024transversallogicalcliffordgates}. A round of mutually commuting CNOTs are performed.]{
\resizebox{0.4\linewidth}{!}{
    \centering     
    \tdplotsetmaincoords{65}{22.5}
    %
}
}
\caption{}
\end{figure}

\begin{figure}[!h]
\centering
\subfloat[\label{fig:CCLP_no_web2}Time slice 2: the `\textit{half-cycle state}' from \cite{chen2024transversallogicalcliffordgates}. ]{
\resizebox{0.4\linewidth}{!}{
    \centering     
    \tdplotsetmaincoords{65}{22.5}
    %
}
}
\qquad
\subfloat[\label{fig:CCLP_no_web3}Time slice 3: the `\textit{post-S state}' from \cite{chen2024transversallogicalcliffordgates}. highly non-local controlled Phase (CZ) gates are performed across the surface code, along with single qubit $S$ and $S^{\dagger}$ gates.]{
\resizebox{0.4\linewidth}{!}{
    \centering     
    \tdplotsetmaincoords{65}{22.5}
    %
}
}
\caption{}
\end{figure}

\begin{figure}[!h]
\centering
\subfloat[\label{fig:CCLP_no_web4}Time slice 4: the `\textit{waning-crescent state}' from \cite{chen2024transversallogicalcliffordgates}.]{
\resizebox{0.4\linewidth}{!}{
    \centering     
    \tdplotsetmaincoords{65}{22.5}
    %
}
}
\caption{}
\end{figure}

\begin{figure}[!h]
\centering
\subfloat[\label{fig:CCLP_no_web6}Time slice 6: the `\textit{end-cycle state}' from \cite{chen2024transversallogicalcliffordgates}. Syndrome qubits are measured and readout in their respective basis.]{
\resizebox{0.4\linewidth}{!}{
    \centering     
    \tdplotsetmaincoords{65}{22.5}
    %
}
}
\qquad
\caption{}
\end{figure}

Figures \ref{fig:CCLP_no_web0} to \ref{fig:CCLP_no_web6} are time slices of the full ZX-diagram of this process. In temporal order, we omit wires entering and exiting all the spiders in the vertical direction for a cleaner representation. 
These ZX-diagrams can be joined in time from figure \ref{fig:CCLP_no_web0} to \ref{fig:CCLP_no_web6} to produce a full timeline ZX-diagram for the CCLP $S$ gate as shown in figure \ref{fig:chen_transversal_S_on_plus_state_X_Y_web}. We shall describe how to obtain the Pauli web describing the logical $Y$ eigenstate initialisation.

\section{Pauli web: logical $X\rightarrow Y$ correlator}
We can now follow the rules from \cite{rodatz2024floquetifyingstabilisercodesdistancepreserving} to generate a Pauli web for the ZX-diagram associated with the CCLP $S$ gate. The Pauli web is a graphical interpretation, connecting the initial to final logical correlators from $X \rightarrow Y$. 

\begin{figure}[!h]
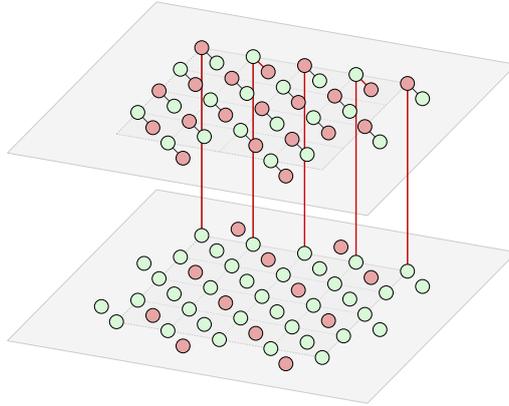

    \centering     
    \tdplotsetmaincoords{65}{22.5}
    \resizebox{0.45\linewidth}{!}{ 

    }
    \caption{The red Pauli web drawn for the first two time slice in the CCLP $S$ gate scheme applied to all $\ket{+}$ initial states.}
    \label{fig:web01}
\end{figure}
\begin{figure}[!h]
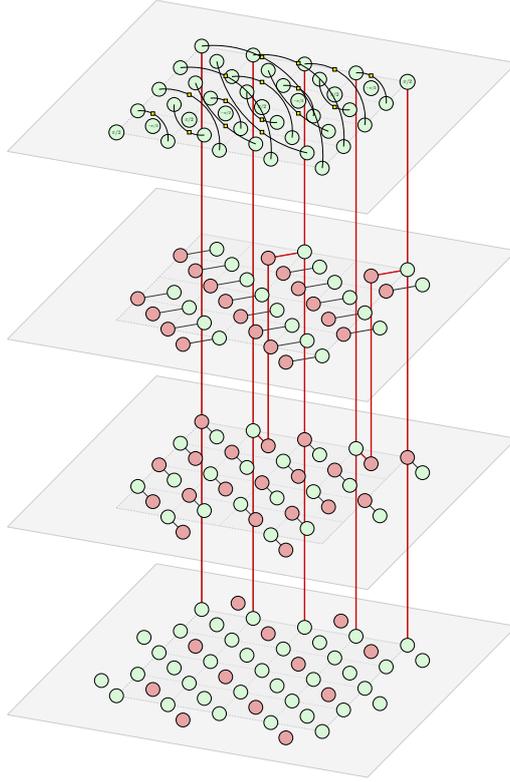

    \centering     
    \tdplotsetmaincoords{65}{22.5}
    \resizebox{0.45\linewidth}{!}{
   
    %
    }
    \caption{\label{fig:web_01234}The red Pauli web drawn up to just before time slice 3.}
\end{figure}

Starting with the first time slices 0 and 1, we can draw the red ($X$) Pauli webs on the qubits located on the top row as shown in figure \ref{fig:web01}. We can then propagate the red Pauli web easily to just before time slice 3 as shown in figure \ref{fig:web_01234}.

\begin{figure}[!h]
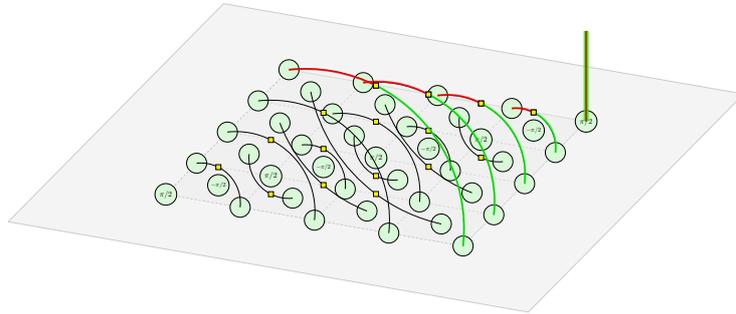

    \centering     
    \tdplotsetmaincoords{65}{22.5}
    \resizebox{0.65\linewidth}{!}{
   

    }
    \caption{\label{fig:timeslice3_zoom_in}The Pauli web on time slice 4, where the CZ gates modify the Pauli web colours leading to the contruction of the logical $Y$ correlator.}
\end{figure}

We shall now focus on time slice 3 and look at the effects of the Pauli web colour changing CZ gates in figure \ref{fig:timeslice3_zoom_in}. The non-local CZ gates linking data and syndrome qubits in a folded way across the counter diagonal of the surface code induces the green ($Z$) Pauli web in the rightmost column of data qubits. The upper right corner $\pi/2$ Z-spider generates the red and green overlapping Pauli web that persists into the future time slices.

\begin{figure}[!h]
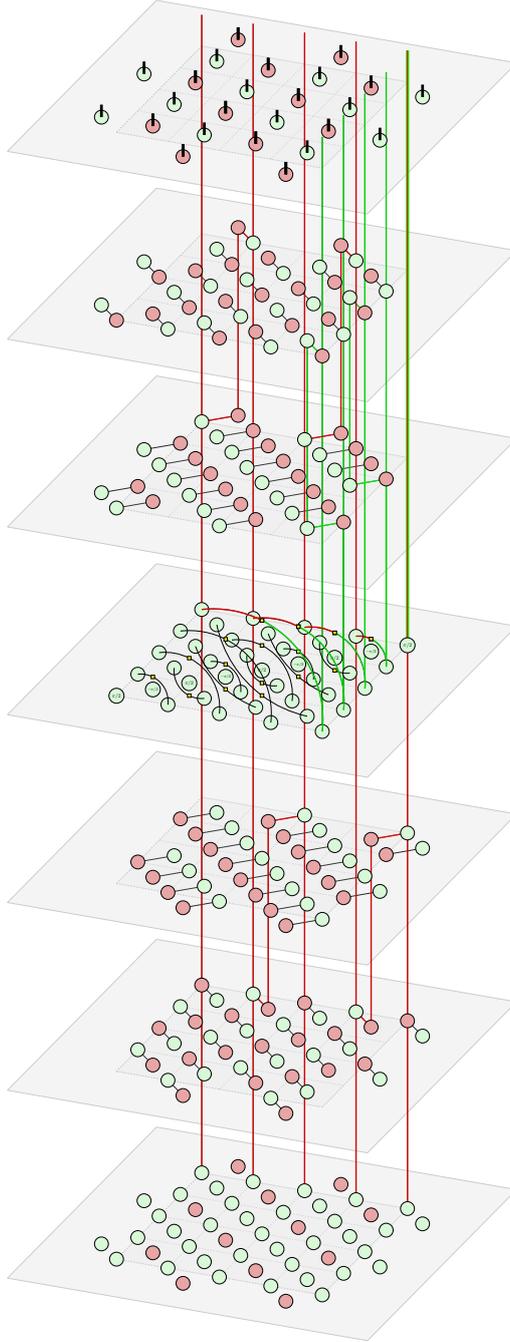

    \centering     
    \tdplotsetmaincoords{65}{22.5}
    \resizebox{0.45\linewidth}{!}{
   
    %
    }
    \caption{\label{fig:chen_transversal_S_on_plus_state_X_Y_web} The Pauli web for the CCLP's fold-transversal $S$ gate applied to the $\ket{+}$ state. Effectively initialising a $\ket{Y}$ state encoded on the surface code. The Pauli web confirms the transformations of logical $X\rightarrow Y$ correlator. Vertical wires in and out of the (normal to the gray planes) spiders in time are omitted for a better presentation.}
\end{figure}

If we propagate the red and green Pauli web into the future time slices as shown in figure \ref{fig:chen_transversal_S_on_plus_state_X_Y_web}, we can recover the logical $Y$ correlator, similar to the calculation performed in \cite{wan2025pauliwebyranglestate}.

\section{Discussion}
In summary, we translated the scheme from \cite{chen2024transversallogicalcliffordgates} to a ZX-diagram and used Pauli web to confirm the action of the CCLP fold-transversal $S$ gate acting on all data qubits initialised in the $\ket{+}$ states. This effectively initialises a logical surface code $\ket{Y}$ state fold-transversally. We will leave the Pauli web verifications of the stabilisers to the readers.

We want to comment that on CCLP's transversal $S$ gate implementation where the authors constructed the fold-transversal gate involving non-local two-qubit gates amongst data and syndrome qubits (separately). As far as we are aware of, this feature is not present in Moussa's \cite{Moussa_2016} or Gidney's \cite{Gidney_Zenodo_Inplace} scheme.

The decoding of syndromes in order to correctly infer the Pauli-frame in transversal gates are challenging \cite{chen2024transversallogicalcliffordgates,wan2024constanttimemagicstatedistillation,wan2024iterativetransversalcnotdecoder,sahay2024errorcorrectiontransversalcnot}. We envisage ZX-calculus and Pauli web can greatly aid the analysis of error propagation in logical operations diagrammatically, potentially devise better ways of decoding logical operations in the future.

\section{Acknowledgements}
Zhenghao Zhong wants to thank Kwok Ho Wan for motivating him to re-do the previous calculation on \cite{Gidney_inplace_2024} again with the recent fold-transversal $S$ gate from \cite{chen2024transversallogicalcliffordgates}. Kwok Ho Wan wishes to thank Zhenghao Zhong for his continued support and for reminding him about the joys of physics. Zhenghao Zhong is currently supported by the ERC Consolidator Grant \# 864828 ``Algebraic Foundations of Supersymmetric Quantum Field Theory'' (SCFTAlg). Zhenghao Zhong acknowledges the TikZ figures in this manuscript are modified version of the TikZ code from \cite{wan2025pauliwebyranglestate}. Kwok Ho Wan proposed the project and outlined the core techniques to study logical non-probabilistic $\ket{Y}$ initialisation. Zhenghao Zhong, re-performed the calculations on the recent scheme \cite{chen2024transversallogicalcliffordgates} and wrote the manuscript.

\bibliography{main}

\begin{thebibliography}{17}
\providecommand{\natexlab}[1]{#1}
\providecommand{\url}[1]{\texttt{#1}}
\expandafter\ifx\csname urlstyle\endcsname\relax
  \providecommand{\doi}[1]{doi: #1}\else
  \providecommand{\doi}{doi: \begingroup \urlstyle{rm}\Url}\fi

\bibitem[Moussa(2016)]{Moussa_2016}
Jonathan~E. Moussa.
\newblock {T}ransversal {C}lifford gates on folded surface codes.
\newblock \emph{Physical Review A}, 94\penalty0 (4), October 2016.
\newblock ISSN 2469-9934.
\newblock \doi{10.1103/physreva.94.042316}.
\newblock URL \url{http://dx.doi.org/10.1103/PhysRevA.94.042316}.

\bibitem[Gidney(2024)]{Gidney_inplace_2024}
Craig Gidney.
\newblock {I}nplace {A}ccess to the {S}urface {C}ode {Y} {B}asis.
\newblock \emph{Quantum}, 8:\penalty0 1310, April 2024.
\newblock ISSN 2521-327X.
\newblock \doi{10.22331/q-2024-04-08-1310}.
\newblock URL \url{http://dx.doi.org/10.22331/q-2024-04-08-1310}.

\bibitem[Gidney(2023)]{gidney2023cleanermagicstateshook}
Craig Gidney.
\newblock Cleaner magic states with hook injection, 2023.
\newblock URL \url{https://arxiv.org/abs/2302.12292}.

\bibitem[Chen et~al.(2024)Chen, Chen, Lu, and Pan]{chen2024transversallogicalcliffordgates}
Zi-Han Chen, Ming-Cheng Chen, Chao-Yang Lu, and Jian-Wei Pan.
\newblock {T}ransversal {L}ogical {C}lifford gates on rotated surface codes with reconfigurable neutral atom arrays, 2024.
\newblock URL \url{https://arxiv.org/abs/2412.01391}.

\bibitem[Fowler et~al.(2012)Fowler, Mariantoni, Martinis, and Cleland]{Fowler_2012}
Austin~G. Fowler, Matteo Mariantoni, John~M. Martinis, and Andrew~N. Cleland.
\newblock {S}urface codes: {T}owards practical large-scale quantum computation.
\newblock \emph{Physical Review A}, 86\penalty0 (3), September 2012.
\newblock ISSN 1094-1622.
\newblock \doi{10.1103/physreva.86.032324}.
\newblock URL \url{http://dx.doi.org/10.1103/PhysRevA.86.032324}.

\bibitem[Li(2015)]{Li_2015}
Ying Li.
\newblock A magic state’s fidelity can be superior to the operations that created it.
\newblock \emph{New Journal of Physics}, 17\penalty0 (2):\penalty0 023037, February 2015.
\newblock ISSN 1367-2630.
\newblock \doi{10.1088/1367-2630/17/2/023037}.
\newblock URL \url{http://dx.doi.org/10.1088/1367-2630/17/2/023037}.

\bibitem[Lao and Criger(2022)]{lao_criger_magic}
Lingling Lao and Ben Criger.
\newblock Magic state injection on the rotated surface code.
\newblock In \emph{Proceedings of the 19th ACM International Conference on Computing Frontiers}, CF '22, page 113–120, New York, NY, USA, 2022. Association for Computing Machinery.
\newblock ISBN 9781450393386.
\newblock \doi{10.1145/3528416.3530237}.
\newblock URL \url{https://doi.org/10.1145/3528416.3530237}.

\bibitem[Gidney(2022)]{Gidney_Zenodo_Inplace}
Craig Gidney.
\newblock Data for "{I}nplace {A}ccess to the {S}urface {C}ode {Y} {B}asis", December 2022.
\newblock URL \url{https://zenodo.org/records/7487893}.

\bibitem[Sahay et~al.(2024)Sahay, Lin, Huang, Brown, and Puri]{sahay2024errorcorrectiontransversalcnot}
Kaavya Sahay, Yingjia Lin, Shilin Huang, Kenneth~R. Brown, and Shruti Puri.
\newblock Error correction of transversal {CNOT} gates for scalable surface code computation, 2024.
\newblock URL \url{https://arxiv.org/abs/2408.01393}.

\bibitem[Wan et~al.(2024)Wan, Webber, Fowler, and Hensinger]{wan2024iterativetransversalcnotdecoder}
Kwok~Ho Wan, Mark Webber, Austin~G. Fowler, and Winfried~K. Hensinger.
\newblock An iterative transversal {CNOT} decoder, 2024.
\newblock URL \url{https://arxiv.org/abs/2407.20976}.

\bibitem[Wan(2024)]{wan2024constanttimemagicstatedistillation}
Kwok~Ho Wan.
\newblock Constant-time magic state distillation, 2024.
\newblock URL \url{https://arxiv.org/abs/2410.17992}.

\bibitem[Coecke and Duncan(2011)]{Coecke_2011}
Bob Coecke and Ross Duncan.
\newblock Interacting quantum observables: categorical algebra and diagrammatics.
\newblock \emph{New Journal of Physics}, 13\penalty0 (4):\penalty0 043016, April 2011.
\newblock ISSN 1367-2630.
\newblock \doi{10.1088/1367-2630/13/4/043016}.
\newblock URL \url{http://dx.doi.org/10.1088/1367-2630/13/4/043016}.

\bibitem[Bombin et~al.(2024)Bombin, Litinski, Nickerson, Pastawski, and Roberts]{Bombin_2024}
Hector Bombin, Daniel Litinski, Naomi Nickerson, Fernando Pastawski, and Sam Roberts.
\newblock Unifying flavors of fault tolerance with the {ZX} calculus.
\newblock \emph{Quantum}, 8:\penalty0 1379, June 2024.
\newblock ISSN 2521-327X.
\newblock \doi{10.22331/q-2024-06-18-1379}.
\newblock URL \url{http://dx.doi.org/10.22331/q-2024-06-18-1379}.

\bibitem[Rodatz et~al.(2024)Rodatz, Poór, and Kissinger]{rodatz2024floquetifyingstabilisercodesdistancepreserving}
Benjamin Rodatz, Boldizsár Poór, and Aleks Kissinger.
\newblock Floquetifying stabiliser codes with distance-preserving rewrites, 2024.
\newblock URL \url{https://arxiv.org/abs/2410.17240}.

\bibitem[Gottesman(1997)]{gottesman1997stabilizercodesquantumerror}
Daniel Gottesman.
\newblock Stabilizer {C}odes and {Q}uantum {E}rror {C}orrection, 1997.
\newblock URL \url{https://arxiv.org/abs/quant-ph/9705052}.

\bibitem[van~de Wetering(2020)]{vandewetering2020zxcalculusworkingquantumcomputer}
John van~de Wetering.
\newblock {ZX}-calculus for the working quantum computer scientist, 2020.
\newblock URL \url{https://arxiv.org/abs/2012.13966}.

\bibitem[Wan and Zhong(2025)]{wan2025pauliwebyranglestate}
Kwok~Ho Wan and Zhenghao Zhong.
\newblock Pauli web of the $|{Y}\rangle$ state surface code injection, 2025.
\newblock URL \url{https://arxiv.org/abs/2501.15566}.

\end{thebibliography}

\end{document}